\documentclass[ALICE,manyauthors]{cernphprep}

\RequirePackage[normalem]{ulem}
\RequirePackage{color}\definecolor{RED}{rgb}{1,0,0}\definecolor{BLUE}{rgb}{0,0,1}

\usepackage[comma,square,numbers,sort&compress]{natbib}
\usepackage{hyperref}
\usepackage{lineno}

\usepackage{graphicx,type1cm,eso-pic,color}
\usepackage{epsfig}
\usepackage{graphicx}
\usepackage{fancyhdr}
\usepackage{pslatex}   
\usepackage{color}
\usepackage{amsmath, amsthm, amssymb} 
\usepackage{endnotes}
\usepackage{xspace}
\usepackage{multirow}
\usepackage{epstopdf}

\begin{document}%

%%%%%%%%%%%%%%%  Title page %%%%%%%%%%%%%%%%%%%%%%%%
\begin{titlepage}
\PHyear{2015}
\PHnumber{019}      % required, will be obtained from PH
\PHdate{30 January}  % required, will be obtained from PH
%

%%% Put your own title + short title here:
\title{Two-pion femtoscopy in p--Pb collisions at $\bf \sqrt{\textit{s}_{\rm NN}}=5.02$ TeV}
\ShortTitle{Two-pion femtoscopy in p--Pb collisions}   % appears on right page headers

%%% Do not change the next lines
\Collaboration{ALICE Collaboration\thanks{See Appendix~\ref{app:collab} for the list of collaboration members}}
\ShortAuthor{ALICE Collaboration} % appears on left page headers, do not change

\begin{abstract}
We report the results of the femtoscopic analysis of pairs of
identical pions measured in p--Pb collisions at
$\sqrt{s_{\mathrm{NN}}}=5.02$~TeV. Femtoscopic radii are determined as a function of
event multiplicity and pair momentum in three spatial dimensions. As
in the pp~collision system, the analysis is complicated by the
presence of sizable background correlation structures in addition to
the femtoscopic signal. The radii increase with event multiplicity and
decrease with pair transverse momentum. When taken at comparable
multiplicity, the radii measured in p--Pb collisions, at high
multiplicity and low pair transverse momentum, are 10--20\% higher
than those observed in pp collisions but below those observed in A--A
collisions. The results are compared to hydrodynamic predictions at
large event multiplicity as well as discussed in the context of calculations based on gluon saturation. 
\end{abstract}
\end{titlepage}
\setcounter{page}{2}

\section{Introduction}
\label{sec:intro}
The Large Hadron Collider (LHC)~\cite{Evans:2008zzb} delivered Pb--Pb collisions at $\sqrt{s_{\rm NN}}=2.76$~TeV in 2010 and in 2011. Several signatures of a quark-gluon plasma were observed, including a strong suppression of high-$p_{\rm T}$~particle production ("jet quenching")~\cite{Aamodt:2010jd,Chatrchyan:2011sx,Aad:2010bu}, as well as collective behavior at low $p_{\rm T}$~\cite{Aamodt:2010pa,ALICE:2011ab} which is well described by hydrodynamic models with a low shear viscosity to entropy ratio. A comparison to reference results from pp~collisions at $\sqrt{s}=0.9$, 2.76, and 7~TeV shows that these effects cannot be described by an incoherent superposition of nucleon-nucleon collisions. As such, they can be interpreted as final-state phenomena, characteristic of the new state of matter~\cite{Adams:2005dq,Adcox:2004mh,Back:2004je,Arsene:2004fa} created in heavy-ion collisions. In order to verify the creation of such a state, p--Pb collisions at $\sqrt{s_{\rm NN}}=5.02$~TeV, where a quark-gluon plasma is not expected to form, were provided by the LHC. In particular, cold nuclear matter effects, such as gluon saturation, which are expected to influence a number of observables, are being investigated~\cite{Salgado:2011wc}.

One of the observables characterizing the bulk collective system is the size of the particle-emitting region at freeze-out which can be extracted with femtoscopic techniques~\cite{Lednicky:2005af,Lisa:2005dd}. In  particular, two-particle correlations of identical pions (referred to as Bose-Einstein, or Hanbury-Brown Twiss "HBT", correlations) provide a detailed picture of the system size and its dependence on the pair transverse momentum and the event multiplicity. Femtoscopy measures the apparent width of the distribution of relative separation of emission points, which is conventionally called the ``radius parameter.'' The radius can be determined independently for three directions: $long$ along the beam axis, $out$ along the pair transverse momentum, and $side$, perpendicular to the other two.  Such measurements were performed at the LHC for central Pb--Pb collisions~\cite{Aamodt:2011mr} as well as for pp collisions at $\sqrt{s}=0.9$ and 7~TeV~\cite{Aamodt:2010jj,Aamodt:2011kd,Khachatryan:2010un,Khachatryan:2011hi}, and compared to results from heavy-ion collisions at lower collision energies. Two clear trends were found: (i) In A--A collisions all radii scale approximately linearly with the cube root of the final state charged particle multiplicity density at midrapidity $\left< \mathrm{d}N_{\rm ch}/\mathrm{d}\eta \right>^{1/3}$ for all three radii separately, consistently with previous findings~\cite{Lisa:2005dd}. For pp collisions, the radii scale linearly with the cube root of charged particle multiplicity density as well, however the slope and intercept of the scaling line are clearly different than for A--A. (ii) A significant, universal decrease of the radii with pair momentum has been observed in A--A collisions, while the analogous trend in pp depends on the considered direction (\emph{out}, \emph{side}, or \emph{long}) and event multiplicity. A transverse momentum dependence of the radii similar to A--A was observed for the asymmetric d--Au collision system at RHIC~\cite{Chajecki:2005qm,Adare:2014vri}. The one-dimensional radii extracted from the ALICE 3-pion cumulant analysis were also investigated in pp, p--Pb, and Pb--Pb collisions at the LHC~\cite{Abelev:2014pja}. For the p--Pb system, at a given multiplicity, the radii were found to be 5--15\% larger than those in pp, while the radii in Pb--Pb were 35--55\% larger than those in p--Pb.

The A--A pion femtoscopy results are interpreted within the hydrodynamic framework as a signature of collective radial flow. Models including this effect are able to reproduce the ALICE data for central collisions~\cite{Bozek:2010wt,Karpenko:2012yf}. Attempts to describe the pp data in the same framework have not been successful so far and it is speculated that additional effects related to the uncertainty principle may play a role in such small systems~\cite{Shapoval:2013jca}. In p--A collisions, hydrodynamic models which assume the creation of a hot and dense system expanding hydrodynamically predict system sizes larger than those observed in pp, and comparable to those observed in lower-energy A--A collisions at the same multiplicity~\cite{Shapoval:2013jca,Bozek:2013df}. However such models have an inherent uncertainty of the initial state shape and size, which can also differ between pp and peripheral A--A collisions.

Alternatively, a model based on gluon saturation suggests that the initial system size in p--A collisions should be similar to that observed in pp collisions, at least in the transverse direction~\cite{Dusling:2013oia,Bzdak:2013zma}. At that stage both systems are treated in the same manner in the Color Glass Condensate model (CGC), so that their subsequent evolution should lead to comparable radius parameter at freeze-out. Ref.~\cite{Schenke:2014zha} suggests a (small) Yang-Mills evolution in addition. The observation of a larger size in the p--A system with respect to pp would mean that a comparable initial state evolves differently in the two cases, which is not easily explained within the CGC approach alone. The d--Au data measured at RHIC suggest that hydrodynamic evolution may be present in such a system, while the ALICE 3-pion analysis at the LHC~\cite{Abelev:2014pja} leaves room for different interpretations. The pion femtoscopic radii as a function of pair transverse momentum from p--Pb collisions at $\sqrt{s_{\rm NN}}=5.02$~TeV, which are reported in this paper, provide additional constraints on the validity of both approaches. 

The paper is organized as follows: In Sec.~\ref{sec:data} the data-taking conditions, together with event and track selections are described. The femtoscopic correlation function analysis, as well as the extraction of the radii and associated systematic uncertainties, and the discussion of the fitting procedure are explained in Sec.~\ref{sec:cfan}. In Sec.~\ref{sec:results} the results for the radii are shown and compared to model predictions. Section~\ref{sec:conclusions} concludes the paper.

\section{Data taking and track reconstruction}
\label{sec:data}

The LHC delivered p--Pb collisions at the beginning of 2013 at $\sqrt{s_{\mathrm{NN}}}=5.02$~TeV (4~TeV and 1.58~TeV per nucleon for the p and Pb beams, respectively). The nucleon--nucleon center-of-mass system is shifted with respect to the ALICE laboratory system by 0.465 unit of rapidity in the direction of the proton beam.

The ALICE detector and its performance are described in Refs.~\cite{Aamodt:2008zz,Abelev:2014ffa}. The main triggering detector is the V0, consisting of two arrays of 32 scintillator counters, which are installed on each side of the interaction point and cover $2.8<\eta_{\rm lab}<5.1$ (V0A, located on the Pb-remnant side), and $-3.7<\eta_{\rm lab}<-1.7$ (V0C). The minimum-bias trigger requires a signal in both V0 detectors within a time window that is consistent with the collision occurring at the center of the ALICE detector. Additionally, specific selection criteria to remove pile-up collisions are applied~\cite{Abelev:2014ffa}. Approximately 80 million minimum-bias events were analyzed.

The analysis was performed in multiplicity classes, which were determined based on the signal from the V0A detector, located along the Pb-going beam. This ensures that the multiplicity determination procedure uses particles at rapidities significantly different from the ones used for the pion correlation analysis, avoiding potential auto-correlation effects. Events are grouped in four multiplicity classes: 0-20\%, 20-40\%, 40-60\%, and 60-90\%, defined  as fractions of the analyzed event sample sorted by decreasing V0A signal, which is proportional to the multiplicity within the acceptance of this detector. Table~\ref{tab:centranges} shows the multiplicity class definitions and the corresponding mean charged-particle multiplicity densities $\left < \mathrm{d}N_{\mathrm{ch}}/\mathrm{d}\eta \right >$ averaged over $|\eta_{\rm lab}|<0.5$ as obtained using the method presented in Ref.~\cite{ALICE:2012xs}. The  $\left < \mathrm{d}N_{\mathrm{ch}}/\mathrm{d}\eta \right >$ values are not corrected for trigger and vertex-reconstruction inefficiency, which is about 4\% for non-single diffractive events~\cite{ALICE:2012xs}.%~\cite{first pPb dN/deta paper}.

\begin{table}[bht!f] \centering

\begin{center}
  \begin{tabular}{c|c}
    \hline
    \multirow{2}{*}{\bf Event class}       & $\left < \mathrm{d}N_{\mathrm{ch}}/\mathrm{d}\eta \right >$ \\ 
                                            & $|\eta_{\rm lab}|<0.5,p_{\mathrm{T}} > 0$~GeV/\emph{c}  \\
    \hline
    %\hline
      60--90\%    & $ 8.2 \pm 0.3 $                 \\
      40--60\%  & $16.1 \pm 0.4$          \\
      20--40\%  & $23.2 \pm 0.5$          \\
       0--20\%   & $35.5 \pm 0.8$          \\
    \hline
  \end{tabular}
  \end{center}
  \caption{
      Definition of the V0A multiplicity classes as fractions of the
      analyzed event sample and their corresponding $\left < \mathrm{d}N_{\mathrm{ch}}/\mathrm{d}\eta(|\eta_{\rm lab}|<0.5,p_{\mathrm{T}}>0) \right >$. The given uncertainties are systematic only since the statistical uncertainties are negligible.} 
  \label{tab:centranges}
\end{table}

Charged track reconstruction is performed using the Time Projection Chamber (TPC) and the Inner Tracking System (ITS). The TPC is a large volume cylindrical gaseous tracking chamber, providing information of particle trajectories and their specific energy loss. The read-out chambers mounted on the endcaps are arranged in 18 sectors on each side (covering the full azimuthal angle) measuring up to 159 samples per track. The TPC covers an acceptance of $|\eta_{\rm lab}| < 0.8$ for tracks which reach the outer radius of the TPC and $|\eta_{\rm lab}| < 1.5$ for shorter tracks. The ITS is composed of position-sensitive silicon detectors. It consists of six cylindrical layers: two layers of Silicon Pixel Detector (SPD) closest to the beam pipe covering $|\eta_{\rm lab}| < 2.0$ and $|\eta_{\rm lab}| < 1.4$ for inner and outer layers respectively, two layers of Silicon Drift Detector in the middle covering $|\eta_{\rm lab}| < 0.9$, and two layers of Silicon Strip Detector on the outside covering $|\eta_{\rm lab}| < 1.0$. The information from the ITS is used for tracking and primary particle selection. The momentum of each track is determined from its curvature in the uniform magnetic field of $0.5$~T oriented along the beam axis, provided by the ALICE solenoidal magnet.   

The primary-vertex position is determined with tracks reconstructed in the ITS and TPC as described in Ref.~\cite{Abelev:2012hxa}. Events are selected if the vertex position along the beam direction is within $\pm 10$~cm of the center of the detector. This ensures a uniform acceptance in $\eta_{\rm lab}$. 

Each track is required to exploit signals in both TPC and ITS. The track segments from both detectors have to match. Additionally, each track is required to have at least one hit in the SPD. A TPC track segment is reconstructed from space points (clusters). Each track is required to be composed of at least 50 out of the 159 such clusters. The parameters of the track are determined by a Kalman fit to the set of TPC+ITS clusters. The quality of the fit $\chi^{2}$ was required to be better than 4 per cluster in the TPC and better than 36 in ITS. Tracks that show a kink topology in the TPC are rejected. To ensure that dominantly primary-particle tracks are selected, the distance of closest approach to the primary vertex is required to be closer than $2.0$~cm in the longitudinal direction and $(0.0105+0.0350\cdot p_{\mathrm{T}}^{-1.1})$~cm, with $p_{\mathrm{T}}$ in GeV/$c$, in the transverse direction. The kinematic range of particles selected for this analysis is $0.12<p_{\mathrm{T}}<4.0$~GeV/$c$ and $|\eta_{\rm lab}|<0.8$. 

The Time-Of-Flight (TOF) detector is used together with the TPC for pion identification. The TOF is a cylindrical detector of modular structure, consisting of 18 azimuthal sectors divided into 5 modules along the beam axis at a radius $\mathrm{r}\simeq380$~cm. The active elements are Multigap Resistive Chambers (MRPC). For both TPC and TOF, the signal (specific energy loss $\mathrm{d}E/\mathrm{d}x$ for the TPC and the time-of-flight for TOF) for each reconstructed particle is compared with the one expected for a pion. The difference is confronted with the detector resolution. The allowed deviations vary between 2 to 5$\sigma$ for the TPC and 2 to 3$\sigma$ for TOF depending on the momentum of the particle. The selection criteria are optimized to obtain a high-purity sample while maximizing efficiency, especially in the regions where the expected signal for other particles (electrons, kaons, and protons for the TPC, kaons for TOF) approaches the pion value. The purity of the pion sample is above 98\%.

The accepted particles from each event are combined to pairs. The two-particle detector acceptance effects, track splitting and track merging, are present. Track splitting occurs when a single trajectory is mistakenly reconstructed as two tracks. ALICE tracking algorithm has been specifically designed to suppress such cases. In a rare event when splitting happens, two tracks are reconstructed mostly from the same clusters in the ALICE TPC. Therefore pairs which share more than 5\% of clusters are removed from the sample. Together with the anti-merging cut described below this eliminates the influence of the split pairs. Track merging can be understood as two-particle correlated efficiency and separation power. In the ALICE TPC, two tracks cannot be distinguished if their trajectories stay close to each other through a significant part of the TPC volume. Although this happens rarely such pairs by definition have low relative momentum and therefore their absence distorts the correlation function in the signal region. Track splitting and track merging are taken into account and corrected with the procedure described in Ref.~\cite{Aamodt:2011kd}. The effect of the two-particle detector acceptance on the final results is similar to what was observed in pp and is limited to low pair relative momentum, where it slightly affects the shape of the correlation function. However, in p--Pb collisions the femtoscopic effect is an order of magnitude wider than any region affected by this inefficiency and, as a consequence, the extracted radii are not affected by the two-track acceptance.

\begin{figure}[tbh!f]
\begin{center}
\includegraphics*[width=8cm]{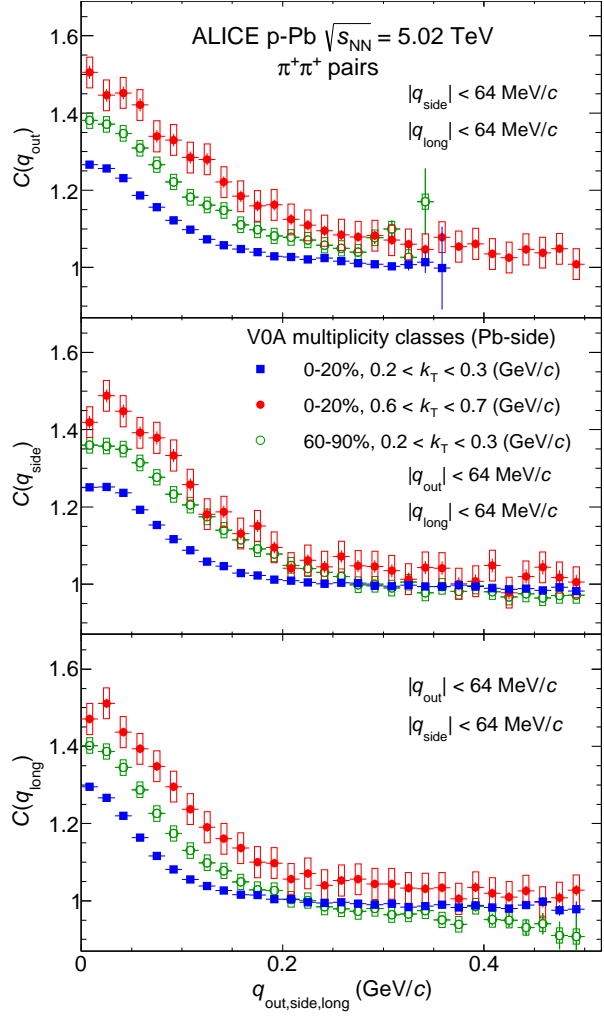} 
\caption[]{(Color online) Projections of the three-dimensional $\pi^+\pi^+$ correlation functions for three selected multiplicity and $k_{\rm T}$ ranges along the $out$ (top), $side$ (middle), and $long$ (bottom) direction. The other components are integrated over the four bins closest to zero in their respective $q$ directions. 
} 
\label{fig:fitex3d}
\end{center}
\end{figure}

\section{Correlation function analysis}
\label{sec:cfan}

\subsection{Construction of the correlation function}

The correlation function $C(\mathbf{p}_{1},\mathbf{p}_{2})$ of two particles with momenta $\mathbf{p}_{1}$ and $\mathbf{p}_{2}$ is defined as
\begin{equation}
C(\mathbf{p}_{1},\mathbf{p}_{2}) = \frac {A(\mathbf{p}_{1},\mathbf{p}_{2})} {B(\mathbf{p}_{1},\mathbf{p}_{2})}.
\label{eq:cfdef}
\end{equation}
The signal distribution $A$ is constructed from pairs of particles from the same event. The background distribution $B$ should be constructed from uncorrelated particles measured with the same single-particle acceptance. It is built using the event mixing method with the two particles coming from two different events for which the vertex positions in beam direction agree within $2$~cm and the multiplicities differ by no more than $1/4$ of the width of the given event class. The denominator is normalized to the number of entries in the numerator, so that the absence of correlation gives a correlation function at unity. 

\begin{figure}[tbh!f]
\begin{center}
\includegraphics*[width=8cm]{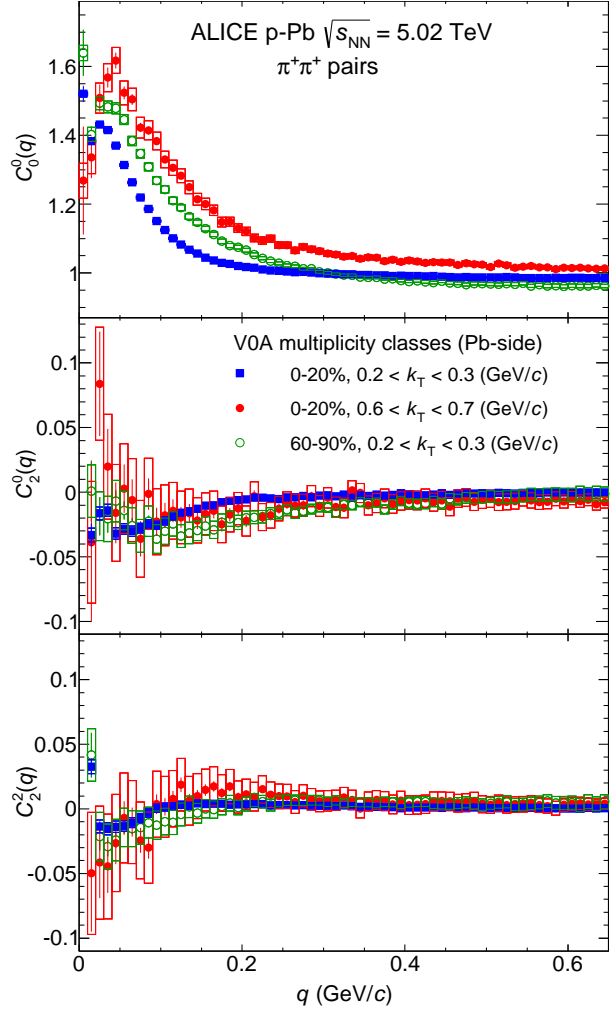}
\caption[]{(Color online) First three non-vanishing components of the SH representation of the $\pi^+\pi^+$ correlation functions for three multiplicity and $k_{\rm T}$ ranges, $l=0$, $m=0$ (top), $l=2$, $m=0$ (middle), and $l=2$, $m=2$ (bottom).} 
\label{fig:fitexsh}
\end{center}
\end{figure}

The femtoscopic correlation is measured as a function of the momentum difference of the pair $\mathbf{q}= \mathbf{p}_2 - \mathbf{p}_1$ as
\begin{equation}
C(\mathbf{q}) = \frac {A(\mathbf{q})} {B(\mathbf{q})},
\label{eq:cfdef}
\end{equation}
where the dependence on the pair total transverse momentum $k_{\mathrm{T}} = 
|\mathbf{p}_{1,\mathrm{T}} + \mathbf{p}_{2,\mathrm{T}}|/2$ is investigated by performing the analysis in the following ranges in $k_{\mathrm{T}}$: 0.2--0.3, 0.3--0.4, 0.4--0.5, 0.5--0.6, 0.6--0.7, 0.7--0.8, and 0.8--1.0~GeV/$c$. The $k_{\mathrm{T}}$~ranges are the same for each multiplicity class resulting in 28 independent correlation functions overall. Systematic uncertainties on the correlation functions are discussed in Sec.~\ref{sec:systunc}.

The momentum difference $\mathbf{q}$ is evaluated in the Longitudinally Co-Moving System (LCMS) frame in which the total longitudinal pair momentum vanishes: $\mathbf{p}_{1,\mathrm{L}} + \mathbf{p}_{2,\mathrm{L}}= 0$, similarly to previous measurements in small systems~\cite{Aamodt:2011kd}. In Fig.~\ref{fig:fitex3d} correlation functions are shown, projected over $128$~MeV/$c$-wide slices along the $q_{\mathrm{out}}$, $q_{\mathrm{side}}$, and $q_{\mathrm{long}}$ axes. An enhancement at low relative momentum is seen in all projections. The width of this correlation peak grows with decreasing multiplicity or with increasing $k_{\rm T}$. The femtoscopic effect is expected to disappear at large $q=|\mathbf{q}|$, with the correlation function approaching unity. We observe, especially for large $k_{\rm T}$ and small multiplicities, that the correlation function is not flat in this region and has different values in different projections\footnote{We note that the overall normalization of the correlation function is a single value for the full three-dimensional object and cannot be independently tuned in all projections.}. The cause may be non-femtoscopic correlations, which are presumably also affecting the shape of the correlation function in the femtoscopic (low $q$) region. This issue  is a major source of systematic uncertainty on the extracted radii and is discussed in detail in Sections~\ref{sec:nonfemto} and \ref{sec:fitting}. 

The pair distributions and the correlation function can be represented in spherical harmonics (SH)~\cite{Chajecki:2009zg,Kisiel:2009iw} alternatively to the traditionally-used Cartesian coordinates. All odd-$l$ and odd-$m$ components of such a representation vanish for symmetry reasons. The important features of the correlation function are fully captured by the following ones: $l=0$, $m=0$ is sensitive to the overall size of the pion source, $l=2$, $m=0$ is sensitive to the difference between the longitudinal and transverse sizes, and $l=2$, $m=2$ reflects the difference between the sidewards and outwards transverse radii. Therefore, three independent sizes  of the source can also be extracted from these three SH components. 

In Fig.~\ref{fig:fitexsh} we show the first three non-vanishing components of the spherical harmonics representation corresponding to the correlation functions  shown in Fig.~\ref{fig:fitex3d}. In the $(0,0)$ component the enhancement at low-$q$ is clearly visible, decreasing (increasing) in width with multiplicity ($k_{\rm T}$). The other two components, $(2,0)$ and $(2,2)$, also show structures in this region, indicating that the source shape is not spherically symmetric in the LCMS frame.

\subsection{Non-femtoscopic structures}
\label{sec:nonfemto}

\begin{figure}[t]
\begin{center}
\includegraphics*[width=8.5cm]{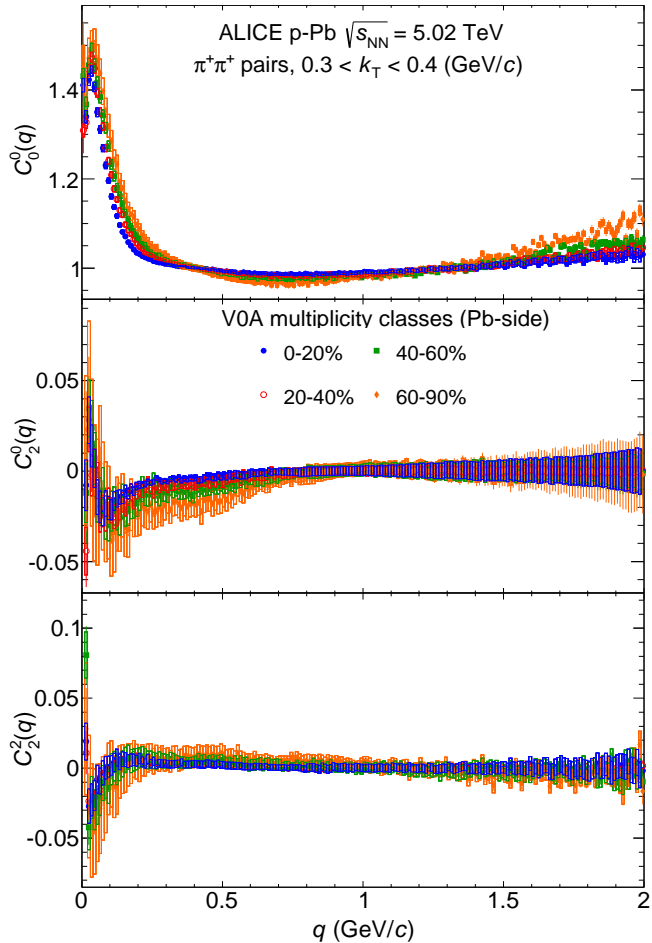} 
\caption[]{(Color online) Dependence of the SH components of the correlation function on event multiplicity in a broad relative momentum range.} 
\label{fig:fitmultbroadsh}
\end{center}
\end{figure}

As mentioned in the discussion of Figs.~\ref{fig:fitex3d} and \ref{fig:fitexsh}, a significant non-femtoscopic correlation is observed in the range in $q$ that is much larger than the characteristic width of the femtoscopic effect. As an example, in Fig.~\ref{fig:fitmultbroadsh} we show
the correlation in the SH representation up to $2.0$~GeV/$c$ in $q$. For the lowest multiplicity, and to a smaller degree at higher multiplicities, a significant slope in low $q$ region is seen in the $(0,0)$ component and a deviation from zero in the $(2,0)$ component up to approximately $1$~GeV/$c$. Similar correlations have been observed by ALICE in pp collisions~\cite{Aamodt:2011kd}. They were interpreted, based on Monte-Carlo model simulations, to be a manifestation of mini-jets, the collimated fragmentation of partons scattered with modest momentum transfer. The lowest multiplicities observed in p--Pb collisions are comparable to those in pp collisions at $\sqrt{s}=7$~TeV. Therefore a similar interpretation of the non-femtoscopic correlations in this analysis is natural. Similar structures have been observed in d--Au collisions by STAR~\cite{Chajecki:2005qm}. This picture is corroborated by  the analysis using the 3-pion cumulants, where expectedly the mini-jet contribution is suppressed~\cite{Abelev:2014pja}.

Two important features of the non-femtoscopic correlation affect the interpretation of our results. First, it is a broad structure, extending up to $1$~GeV/$c$ and we have to assume that it also extends to 0~GeV/$c$ in $q$. Therefore it affects the extracted femtoscopic radii and has to be taken into account in the fitting procedure. It can be quantified in the large $q$ region and then extrapolated, with some assumptions, to the low $q$ region, under the femtoscopic peak. The procedure leads to a systematic uncertainty. Secondly, it becomes visibly larger as multiplicity decreases and also as $k_{\rm T}$~increases, which is consistent with the mini-jet-like correlation.

\begin{figure}[t]
\begin{center}
\includegraphics*[width=8cm]{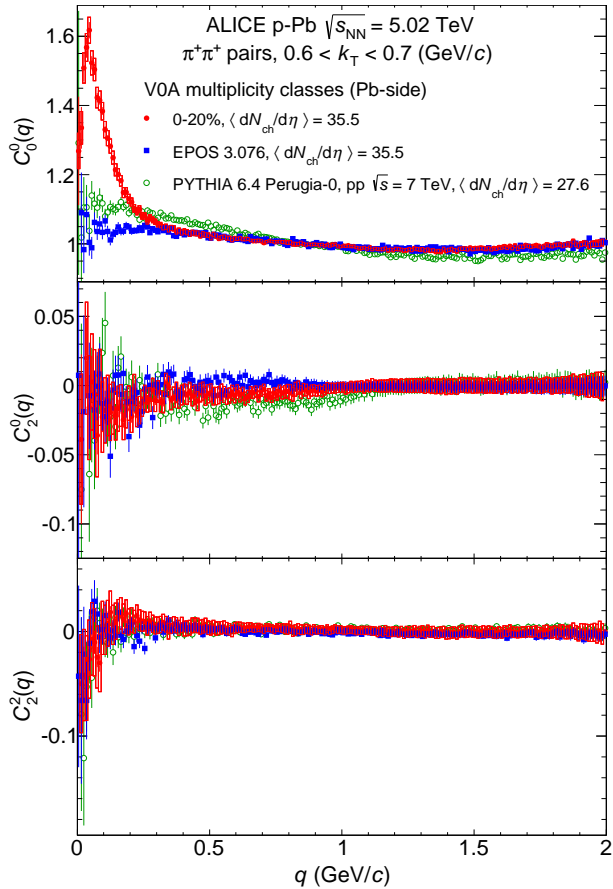}
\caption[]{(Color online) First three non-vanishing components of the SH representation of the $\pi^+\pi^+$ correlation functions for a selected multiplicity and $k_{\rm T}$ range, compared to a calculation from EPOS 3.076 \cite{Werner:2013tya,Werner:2013ipa} (generator level only) and PYTHIA 6.4 Perugia-0 tune \cite{Sjostrand:2006za,Skands:2010ak} for pp at $\sqrt{s}=7$~TeV (full simulation with detector response).
} 
\label{fig:datavsmodels}
\end{center}
\end{figure}

The background was studied in Monte Carlo models, such as, for p--Pb collisions, AMPT \cite{Lin:2004en}, HIJING \cite{Wang:1991hta}, DPMJET \cite{Roesler:2000he}, EPOS 3.076~\cite{Werner:2013tya,Werner:2013ipa}, and PYTHIA 6.4 (Perugia-0 tune)~\cite{Sjostrand:2006za,Skands:2010ak} for pp~collisions at similar multiplicities. In all cases the Monte Carlo correlation functions exhibit significant structures similar to the long-range effects observed in data, which is another argument for their non-femtoscopic origin. However, quantitative differences in the magnitude and shape of these structures when compared to those observed in data are seen for AMPT, HIJING and DPMJET. These models are therefore unsuitable for a precise characterization of the background, which is needed for the fitting procedure. The only models that qualitatively describe the features of the background (enhancement at low $q$, growing with $k_{\rm T}$, and falling with multiplicity) are EPOS 3.076 and PYTHIA 6.4 (Perugia-0 tune), which was also used in the pp~analysis~\cite{Aamodt:2011kd}. We note that PYTHIA simulation included full detector response modelling, while the EPOS 3.076 one did not. The comparison with data is shown in Fig.~\ref{fig:datavsmodels}. The behavior of the correlation is well reproduced above 0.5 GeV/$c$ in $q$, where non-femtoscopic correlations are expected to dominate. At low $q$, below 0.3~GeV/$c$, the data and models diverge, which is expected, as the femtoscopic correlations are not included in the model calculation. Above 0.3~GeV/$c$, EPOS reproduces the $(0,0)$ component well, PYTHIA slightly overestimates the data. For the $(2,0)$ and $(2,2)$ components, which describe the three-dimensional shape of the non-femtoscopic correlations, PYTHIA is closer to the experimental data. Overall, for like-sign pairs, both models are reasonable approximations of the non-femtoscopic background. We use these models to fix the background parameters in the fitting procedure.  

Similarly to the pp analysis~\cite{Aamodt:2011kd}, the unlike-sign pairs have also been studied. We found that correlations in the $(0,0)$ component of PYTHIA are slightly larger than in data in the femtoscopic region for all $k_{\rm T}$ ranges, and similar to data in the $(2,0)$ and $(2,2)$ components. EPOS was found to reasonably describe the unlike-sign pairs for low $k_{\rm T}$ ranges and has smaller correlation than data in the $(0,0)$ component in higher $k_{\rm T}$ ranges.

\subsection{Fitting the correlation functions}
\label{sec:fitting}

The space-time characteristics of the source are reflected in the correlation function
\begin{equation}
C(\mathbf{q}) = \int S(r, \mathbf{q}) |\Psi(\mathbf{q},r)|^{2}
d^{4}r,
\label{eq:koonpratt}
\end{equation}
where $r$ is the pair space-time separation four-vector. $S$ is the source emission function, interpreted as a probability density describing the emission of a pair of particles with a given relative momentum and space-time separation. $\Psi$ is the two-particle interaction kernel. 

Previous femtoscopy studies in heavy-ion collisions at SPS~\cite{Adamova:2002wi}, 
RHIC~\cite{Adler:2001zd,Adams:2003ra,Adams:2004yc,Abelev:2009tp,Adcox:2002uc,Adler:2006as,Afanasiev:2007kk} and at the LHC~\cite{Aamodt:2011mr} used a Gaussian static source $S$
\begin{equation}
S(r) \approx \exp \left(-\frac{r^{2}_{\mathrm{out}}}
    {4{R^{\mathrm{G}}_{\mathrm{out}}}^2} - \frac{r^{2}_{\mathrm{side}}}
    {4{R^{\mathrm{G}}_{\mathrm{side}}}^2} - \frac{r^{2}_{\mathrm{long}}}
    {4{R^{\mathrm{G}}_{\mathrm{long}}}^2} 
 \right ).
\label{eq:sfun}
\end{equation}
The $R^{\mathrm{G}}_{\mathrm{out}}$, $R^{\mathrm{G}}_{\mathrm{side}}$, and $R^{\mathrm{G}}_{\mathrm{long}}$ parameters describe the single-particle source size in the LCMS in the  $out$, $side$, and $long$ directions, respectively.

The Gaussian source provides a commonly used approximation of the source size and was used to compare to other experimental results, especially the ones coming from A--A collisions, where the source shape is more Gaussian than in small systems. While pursuing the standard procedure with the Gaussian assumption, we also carefully look for any deviations between the fit function and data which might suggest a significantly non-Gaussian shape of the source, which would be an important similarity to the pp~case.

In the analysis of pp~collisions by ALICE~\cite{Aamodt:2011kd}, a Gaussian is used together with other source shapes, exponential and Lorentzian~\cite{Aamodt:2011kd}. A Lorentzian parametrization in the $out$ and $long$ directions and a Gaussian parametrization in the $side$ direction were found to fit the data best according to $\chi^2/\mathrm{ndf}$. Therefore, we use this source parametrization also in the analysis of p--Pb collisions
\begin{equation}
S(r) \approx
\frac{1}
    {r^{2}_{\mathrm{out}}+{R^{\mathrm{E}}_{\mathrm{out}}}^{2}} 
\exp \left(- \frac{r^{2}_{\mathrm{side}}} {4{R^{\mathrm{G}}_{\mathrm{side}}}^{2}} \right)
\frac{1} {r^{2}_{\mathrm{long}}+{R^{\mathrm{E}}_{\mathrm{long}}}^{2}} .
\label{eq:sfunege}
\end{equation}
The corresponding source sizes in $out$ and $long$ are $R^{\mathrm{E}}_{\mathrm{out}}$ and $R^{\mathrm{E}}_{\mathrm{long}}$, while for the $side$ direction the size parameter $R^{\mathrm{G}}_{\mathrm{side}}$ is the same as in the Gaussian case.

For identical pions, which are bosons, $\Psi$ must be symmetrized. Since charged pions also interact via the Coulomb and strong Final State Interactions (FSI), $|\Psi|^{2}$ corresponds to the Bethe-Salpeter amplitude~\cite{Lednicky:2005tb}. For like-sign pion pairs the contribution of the strong interaction is small for the expected source sizes \cite{Lednicky:2005tb}, and is neglected here. The used $\Psi$ therefore is a symmetrized Coulomb wave function. It is approximated by separating the Coulomb part and integrating it separately, following the procedure
known as Bowler-Sinyukov fitting~\cite{Bowler:1991vx,Sinyukov:1998fc}, which was used previously for larger sizes observed in Pb--Pb~\cite{Aamodt:2011mr} as well as smaller sizes observed in pp~collisions~\cite{Aamodt:2011kd}. In this approximation the integration of Eq.~(\ref{eq:koonpratt}) with $S$ given by Eq.~(\ref{eq:sfun}) results in the following functional form for the correlation function which is used to fit the data
\begin{eqnarray}
\label{eq:cfun}
C_{\mathrm{f}}(\mathbf{q}) &=&  (1-\lambda) \\ 
&+& \lambda K_{\mathrm{C}}(q) \left [ 1 +
  \exp(-{R^{\rm G}_{\mathrm{out}}}^{2}q_{\mathrm{out}}^{2}-{R^{\rm G}_{\mathrm{side}}}^{2}q_{\mathrm{side}}^{2}-{R^{\rm G}_{\mathrm{long}}}^{2}q_{\mathrm{long}}^{2})  
\right ].\nonumber
\end{eqnarray}
The function $K_{\mathrm{C}}(q)$ is the Coulomb part of the two-pion wave function integrated over the spherical Gaussian source with a fixed radius. The value of this radius is chosen to be $2$~fm. Its uncertainty has systematic effects on the final results (see Sec.~\ref{sec:systunc}). This form of the correlation function from Eq.~(\ref{eq:cfun}) is denoted in the following as GGG. Similarly for the source shape given by Eq.~\eqref{eq:sfunege} the correlation function is
\begin{eqnarray}
\label{eq:cfunege}
C_{\mathrm{f}}(\mathbf{q})& =& (1-\lambda) \\ 
&+& \lambda K_{\mathrm{C}} \left [ 1 +
  \exp \left (-\sqrt{{R^{\mathrm{E}}_{\mathrm{out}}}^{2}q^{2}_{\mathrm{out}}}-{R^{\mathrm{G}}_{\mathrm{side}}}^{2}q_{\mathrm{side}}^{2}-\sqrt{{R^{\mathrm{E}}_{\mathrm{long}}}^{2}q^{2}_{\mathrm{long}}} \right )  
\right ]. \nonumber
\end{eqnarray}
It has an exponential shape in $out$ and $long$ and a Gaussian shape in the $side$ direction. Therefore it is referred to as EGE form of the correlation function. Parameter $\lambda$ in Eqs.~(\ref{eq:cfun}) and (\ref{eq:cfunege}) represents the correlation strength.
 
Additionally, a component $\Omega$ describing non-femtoscopic correlations needs to be introduced. There is no \emph{a priori} functional form which can be used for this component. Several of its features can be deduced from the correlations shown in Figs.~\ref{fig:fitex3d}, \ref{fig:fitexsh}, and
\ref{fig:fitmultbroadsh}: it has to allow for different shapes in the $out$, $side$, and $long$ directions; in the $(0,0)$ component it has to extrapolate smoothly to low $q$ and have a vanishing slope at $q=0$. Since this structure is not known, maximum information about its shape and magnitude should be gained from an observation of the raw correlation functions and the corresponding effects in Monte Carlo, in as many representations as possible. It is therefore crucial to simultaneously use the Cartesian and SH representations as they provide complementary ways to study the correlation shape.

An ad-hoc, Monte-Carlo driven parametrization of the non-femto background that reasonably describes the correlation function is $\Omega$, composed of three independent 1-dimensional functions $\Omega_0^0$ (Gaussian plus fixed constant), $\Omega_2^0$ (Gaussian plus variable constant) and $\Omega_2^2$ (Gaussian plus an additional linear component)
\begin{eqnarray}
\Omega_{0}^{0}(q) &=& N_{0}^{0} \left [1+\alpha_{0}^{0}\exp \left (-\frac{q^2}{2(\sigma_0^0)^2} \right ) \right ],\\
\Omega_{2}^{0}(q) &=& \alpha_{2}^{0}\cdot\exp \left (-\frac{q^2}{2(\sigma_2^0)^2} \right )+\beta_2^0 ,\\
\Omega_{2}^{2}(q) &=& \alpha_{2}^{2}\cdot\exp \left (-\frac{q^2}{2(\sigma_2^2)^2} \right ) +\beta_2^2+\gamma_2^2\cdot q,
\end{eqnarray}
where $N_0^0$, $\alpha_0^0$, $\sigma_0^0$, $\alpha_2^0$, $\sigma_2^0$, $\alpha_2^2$,
$\sigma_2^2$, and $\gamma_2^2$ are fixed to the values obtained from fits to Monte Carlo events, separately for each multiplicity and $k_{\rm T}$ range. In the fit procedure the $\beta_2^0$ and $\beta_2^2$ parameters are kept free. This results in the following fit formula
\begin{equation}
\begin{split}
C(\mathbf{q}) = N\cdot C_{\mathrm{f}}(\mathbf{q}) \cdot \left [ \Omega_0^0(q) \cdot Y^0_0(\theta,\varphi)+\Omega_2^0(q) \cdot Y_2^0(\theta,\varphi)+ \Omega_2^2(q) \cdot Y_2^2(\theta,\varphi) \right ],
\end{split}
\label{eq:C3D}
\end{equation}
where $N$ is the overall normalization factor and $Y_0^0(\theta,\varphi)$, $Y_2^0(\theta,\varphi)$, and $Y_2^2(\theta,\varphi)$ are the real parts of the relevant spherical harmonic functions. 

The fit is performed with the log-likelihood method in three dimensions for the Cartesian representation. 
The Gaussian fit reproduces the overall width of the femtoscopic correlation in all cases. The background component
describes the behavior of the correlation at large $q$, but can also have non-zero correlation at 0 in $q$. 

\begin{figure}[!t]
\begin{center}
\includegraphics*[width=8cm]{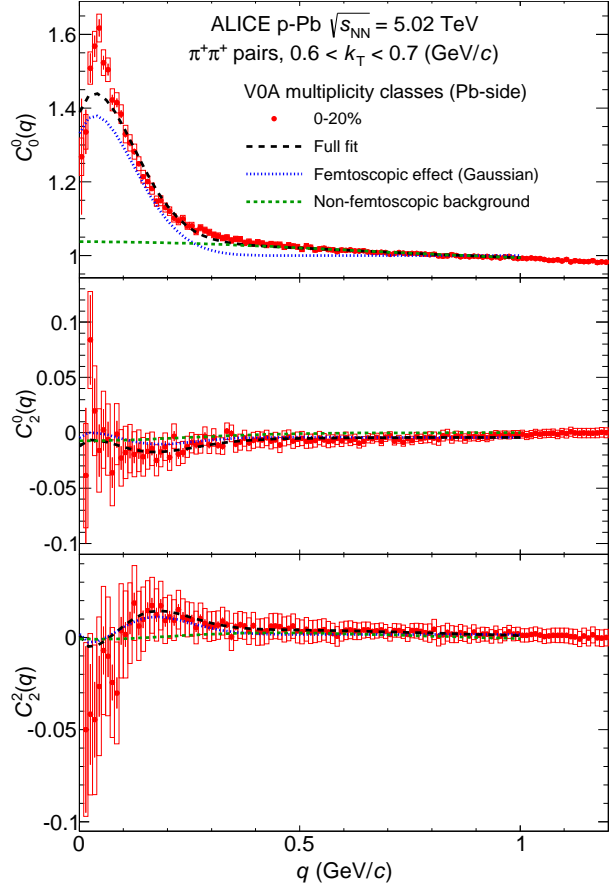} 
\caption[]{(Color online) First three non-vanishing components of the SH representation of the $\pi^+\pi^+$ correlation functions for three multiplicity and $k_{\rm T}$ combinations, $l=0$, $m=0$ (top), $l=2$, $m=0$ (middle), and $l=2$, $m=2$ (bottom). The lines show the corresponding components of the Gaussian (GGG) fit.} 
\label{fig:funSH_fit}
\end{center}
\end{figure}

\begin{figure}[!t]
\begin{center}
\includegraphics*[width=8cm]{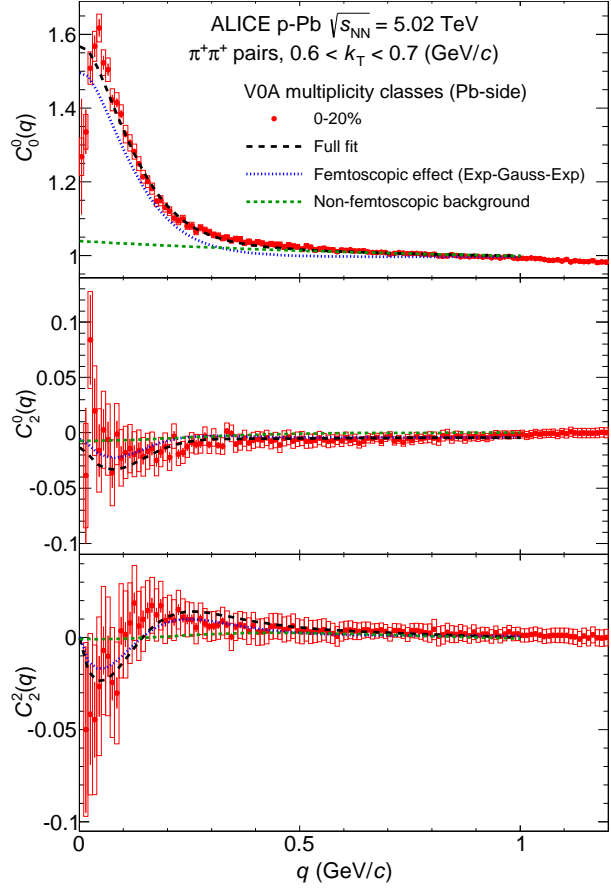}  
\caption[]{(Color online) First three non-vanishing components of the SH representation of the $\pi^+\pi^+$ correlation functions for three multiplicity and $k_{\rm T}$ combinations, $l=0$, $m=0$ (top), $l=2$, $m=0$ (middle), and $l=2$, $m=2$ (bottom). The lines show the corresponding components of the EGE fit.} 
\label{fig:funSH_fitEGE}
\end{center}
\end{figure}

A corresponding fit is also performed for the SH representation of the correlation, which is shown in Fig.~\ref{fig:funSH_fit}. The formula from Eq.~(\ref{eq:cfun}) or Eq.~(\ref{eq:cfunege}) (for the GGG fit or the EGE fit, respectively) is numerically integrated on a $\varphi-\theta$ sphere for each $q$ bin, with proper $Y_{l}^{m}$ weights, to produce the three components of the SH decomposition. Statistical uncertainties on each component as well as the covariance matrix between them are taken into account in this simultaneous fit to the three histograms. The results are shown in Fig.~\ref{fig:funSH_fit}. The fit describes the general direction-averaged width of the correlation  function, shown in the upper panel. The background component $\Omega$ describes the behavior at large $q$ but also contributes to the correlation at low $q$. The shape in three-dimensional space, captured by the $(2,0)$ and $(2,2)$ components, is also a combination of the
femtoscopic and non-femtoscopic correlations.

Overall the GGG fit describes the width of the correlation but the data at low $q$ are not perfectly reproduced, which can be attributed to the limitations of the Bowler-Sinyukov formula as well as to the non-Gaussian, long-range tails which are possibly present in the source. Some deviations from the pure Gaussian shape can also be seen for the $long$ direction for the higher multiplicities. The EGE fit (Eq.~(\ref{eq:cfunege})) better reproduces the correlation peak in the (0,0) component, as shown in Fig.~\ref{fig:funSH_fitEGE}. The (2,0) and (2,2) components show similar quality of the fit. The $\chi^{2}$ values for both fits are comparable.

\subsection{Systematic uncertainties on the radii}
\label{sec:systunc}

\begin{table}[tb]

\begin{center}
\begin{tabular}{l|c|c|c}
\hline
{\bf Uncertainty source} & ${\it R}^{\rm G}_{\rm out}$ (\%) & ${\it R}^{\rm G}_{\rm side}$ (\%) &
${\it R}^{G}_{\rm long}$ (\%)  \\
\hline
CF representation \&  & \multirow{2}{*}{5--32}& \multirow{2}{*}{4--22}& \multirow{2}{*}{4--35} \\
Background parametrization & & &\\
Fit-range dependence & 10 & 8 & 10 \\
$\pi^{+}\pi^{+}$ vs. $\pi^{-}\pi^{-}$ & 3 & 3 & 3 \\
Momentum resolution correction & 3 & 3 & 3 \\
Two-track cut variation & $<1$ & $<1$ & $<1$ \\
Coulomb correction & $<1$ & $<1$ & $<1$ \\
\hline
{\bf Total correlated} & 12--34 & 9--24 & 11--36 \\
{\bf Total} & 12--34 & 11--24 & 12--36 \\
\hline
\end{tabular}
\end{center}
\caption{List of contributions to the systematic uncertainty of the femtoscopic radii extracted via GGG fits. Values are averaged over $k_{\rm T}$ and multiplicity except for the first row where a minimum--maximum range is shown. } 
\label{tab:systerror}
\end{table}

\begin{table}[tb]
\begin{center}
\begin{tabular}{l|c|c|c}
\hline
{\bf Uncertainty source} & ${\it R}^{E}_{\rm out}$ (\%) & ${\it R}^{\rm G}_{\rm side}$ (\%) &
${\it R}^{\rm E}_{\rm long}$ (\%)  \\
\hline
CF representation \&  & \multirow{2}{*}{4--18}& \multirow{2}{*}{3--14}& \multirow{2}{*}{8--20} \\
Background parametrization & & &\\
Fit-range dependence & 10 & 6 & 10 \\
$\pi^{+}\pi^{+}$ vs. $\pi^{-}\pi^{-}$ & 3 & 3 & 3 \\
Momentum resolution correction & 3 & 3 & 3 \\
Two-track cut variation & $<1$ & $<1$ & $<1$ \\
Coulomb correction & $<1$ & $<1$ & $<1$ \\
\hline
{\bf Total correlated} & 11--21 & 7--16 & 13--23 \\
{\bf Total} & 12--21 & 8--16 & 14--23 \\
\hline
\end{tabular}
\end{center}
\caption{List of contributions to the systematic uncertainty of the femtoscopic radii extracted via EGE fits. Values are averaged over $k_{\rm T}$ and multiplicity except for the first row where a minimum--maximum range is shown. } 
\label{tab:systerrorEGE}
\end{table}
The analysis was performed separately for positively and negatively charged pions. For the practically zero-net-baryon-density system produced at the LHC they are expected to give consistent results. Both datasets are statistically consistent at the correlation function level.

The main contributions to the systematic uncertainty are given in Table~\ref{tab:systerror} for GGG radii and in Table~\ref{tab:systerrorEGE} for EGE radii.

We used two alternative representations (Cartesian and spherical harmonics) of the correlation function. The same functional form for both of them was used for the fitting procedure. However, the implementation of the fitting procedure is quite different: log-likelihood for Cartesian vs. regular $\chi^{2}$ fit for SH, 3D Cartesian histogram vs. three 1D histograms, cubic or spherical fitting range in the $(q_{\rm out},q_{\rm side},q_{\rm long})$ space. Therefore the fits to the two representations may react in a systematically different way to the variation of the fitting procedure (fit ranges, Bowler-Sinyukov approximation, etc.). 

The fitting procedure requires the knowledge of the non-femtoscopic background shape and magnitude. Two models were used to estimate it, EPOS 3.076~\cite{Werner:2013ipa} and PYTHIA 6.4 (Perugia-0 tune)~\cite{Sjostrand:2006za,Skands:2010ak}, as described in Sec.~\ref{sec:nonfemto}.

In addition the correlation function shape is not ideally described by a Gaussian form. The EGE form is better (lower $\chi^2$ values for the fit), but still not exactly accurate. As a result the fit values depend on the fitting range used in the procedure of radius extraction. We have performed fits with an upper limit of the fit range varied between 0.3~GeV/$c$ and 1.1~GeV/$c$.

The three effects mentioned above are the main sources of systematic uncertainty on the radii. Their influence, averaged over the event multiplicity and pair $k_{\rm T}$, is given in Table~\ref{tab:systerror} and Table~\ref{tab:systerrorEGE}.  The background parametrization and the CF representation effects lead to systematic uncertainties less than 10\% at low $k_{\mathrm{T}}$ and up to 35\% for large $k_{\mathrm{T}}$ and low multiplicities. In particular, the radius could not be reliably extracted for the two highest $k_{\mathrm{T}}$ ranges in the lowest multiplicity range, therefore these two sets of radii are not shown. Moreover, radii obtained with the background parametrization from PYTHIA are always larger than the ones obtained with the EPOS parametrization. These uncertainties are correlated between $k_{\rm T}$ ranges. Similarly, the radii from the narrow fit range are always on average 10\% higher than the ones from the wide fit range. This also gives a correlated contribution to the systematic uncertainty. The final radii are calculated as an average of four sets of radii -- the two representations with both EPOS and PYTHIA background parametrization. The systematic uncertainties are symmetric and equal to the largest difference between the radius and one of the four sets of radii.
  
The effect of the momentum resolution on the correlation function was studied using a Monte Carlo simulation. For tracks with a low $p_{\rm T}$, below 1~$\mathrm{GeV}/c$, the momentum resolution in the TPC is better than 1\%. Smearing of the single particle momenta reduces the height and increases the width of the correlation function. It was estimated that this effect changes the reconstructed radius by 2\% for a system size of 2 fm and 3\% for a size of 3~fm. Therefore, the 3\% correlated contribution from momentum resolution is always added to the final systematic uncertainty estimation.

Smaller sources of systematic uncertainties include those originating from the difference between positively and negatively charged pion pairs (around 3\%), track selection variation (less than 1\%) and from the Coulomb factor  (less than 1\%). All the uncertainties are added in quadrature.

\begin{figure}[tbh!f]
\begin{center}
\includegraphics*[width=8cm]{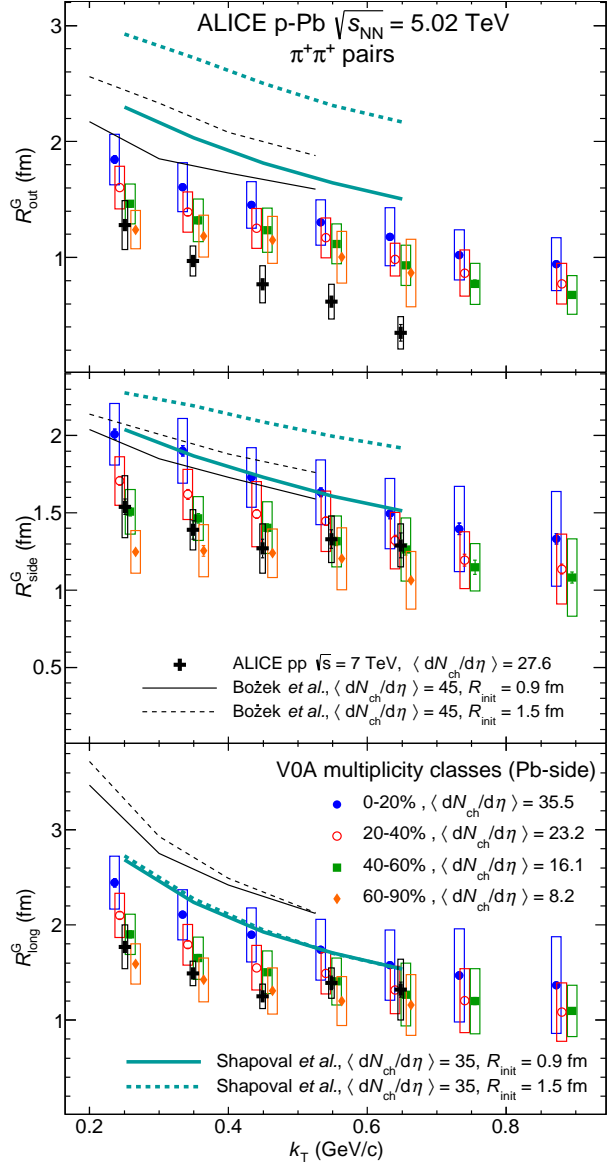} 
\caption[]{(Color online) Femtoscopic radii (GGG fit) as a function of the pair transverse momentum $k_{\rm T}$ for four multiplicity classes. For comparison, radii from high-multiplicity pp~collisions at $\sqrt{s}=7$~TeV~\cite{Aamodt:2010jj} and 4 predictions for p--Pb~\cite{Bozek:2013df,Shapoval:2013jca} are shown as crosses and lines, respectively. Top, middle, and bottom panel show $R_{\mathrm{out}}$, $R_{\mathrm{side}}$, and $R_{\mathrm{long}}$ radii, respectively. The points for multiplicity classes 20-40\% and 40-60\% have been slightly shifted in $k_{\rm T}$ for visibility.}
\label{fig:pbktcentradii}
\end{center}
\end{figure}

\section{Results}
\label{sec:results}

\subsection {Three-dimensional radii}
We have extracted $R_{\rm out}$, $R_{\rm side}$, and $R_{\rm long}$ in intervals of multiplicity and $k_{\rm T}$, which results in 26 radii in each direction. The fit procedure did not allow to reliably extract values of the radii for the two highest $k_{\rm T}$ ranges in the 60--90\% multiplicity class. For the GGG fit, they are shown in Fig.~\ref{fig:pbktcentradii}. The radii are in the range of 0.6 to 2.4 fm in all directions and universally decrease with $k_{\rm T}$. The magnitude of this decrease is similar for all multiplicities in the $out$ and $long$ directions, and is visibly increasing with multiplicity in the $side$ direction. The radii rise with event multiplicity. The plot also shows data from pp collisions at $\sqrt{s}=7$~TeV~\cite{Aamodt:2011kd} at the highest multiplicities measured by ALICE, which is slightly higher than the multiplicity measured for the 20-40\% V0A signal range in the p--Pb analysis. At small $k_{\rm T}$ the pp radii are lower by 10\% (for $side$) to 20\% (for $out$) than the p--Pb radii at the same $\left< \mathrm{d}N_{\rm ch}/\mathrm{d}\eta \right>^{1/3}$. At high $k_{\rm T}$ the difference in radius grows for $R_{\rm out}$, while for $R_{\rm long}$ the radii for both systems become comparable. The distinct decrease of radii with $k_{\rm T}$ is observed both in pp and p--Pb.

The correlation strength $\lambda$ increases with $k_{\rm T}$ from 0.44 to 0.58 for the collisions with highest multiplicities. It is also higher for low multiplicity collisions, with a difference of 0.1 between collisions with highest multiplicities and lowest multiplicities. A non-constant $\lambda$ parameter as a function of $k_{\rm T}$ is an indication of a non-Gaussian shape of the correlation function. The correlation functions are normalized to the ratio of the number of pairs in the signal and background histograms. The positive correlation at low $q$ has to be then compensated by the normalization parameter $N$, which is in the range of 0.9--1.0. The $\chi^2/\mathrm{ndf}$ for the three-dimensional fit is on the order of 1.2. 

The extracted background parameters indicate that this contribution increases with $k_{\rm T}$ and decreases with multiplicity, which is consistent with qualitative expectations for the mini-jet effect. The shape of the background is not spherical, leading to finite contributions to the $(2,0)$
and $(2,2)$ components. The constant shift in these components, given by $\beta^0_2$ and $\beta^2_2$ respectively, is only significant for the $(2,0)$ component in lower multiplicities.

\begin{figure}[!ht]
\begin{center}
\includegraphics*[width=8cm]{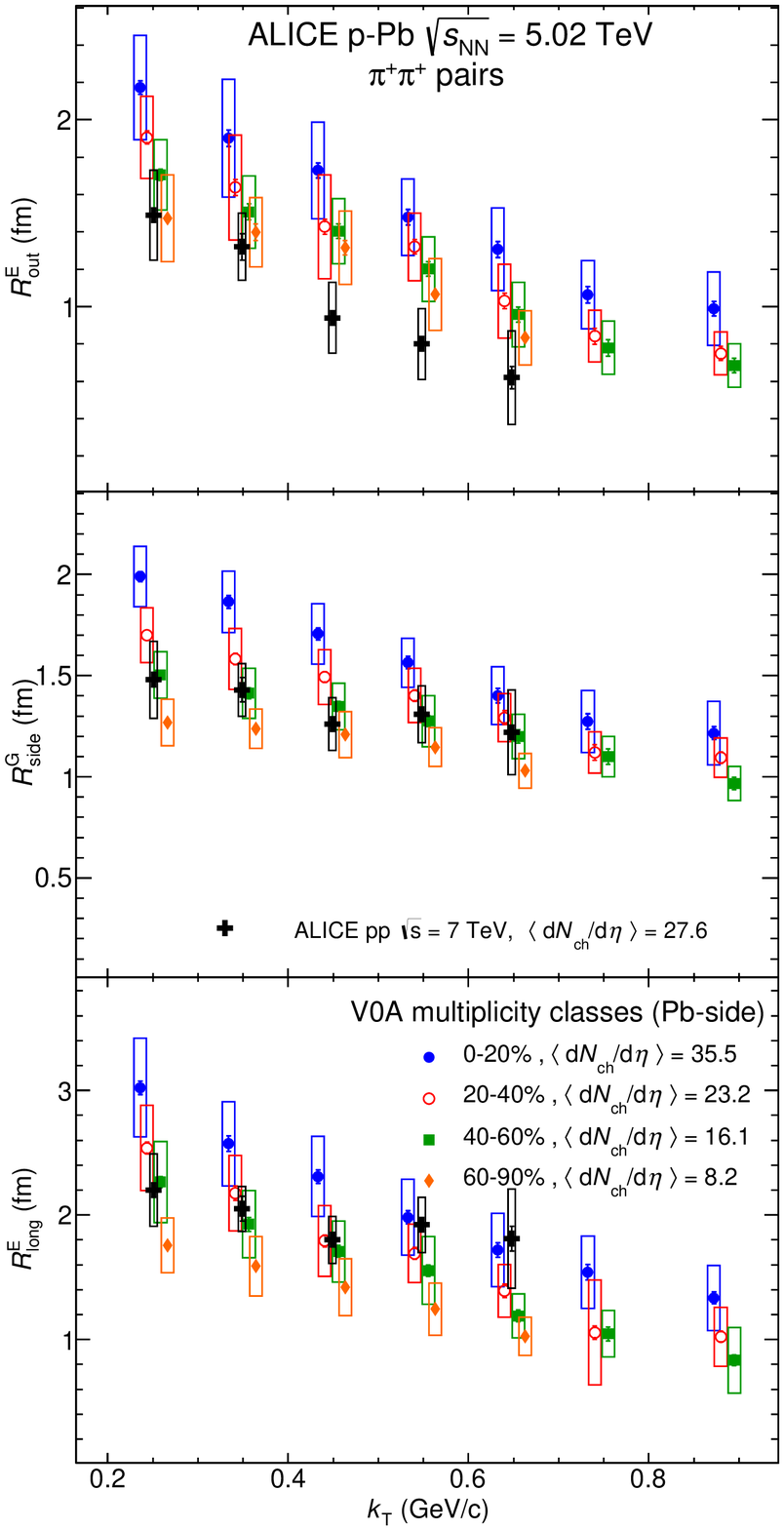} 
\caption[]{(Color online) Femtoscopic radii (EGE fit) as a function of the pair transverse momentum $k_{\rm T}$ for four multiplicity classes. For comparison, radii from high-multiplicity pp~collisions at $\sqrt{s}=7$~TeV~\cite{Aamodt:2010jj} are shown. Top,  middle, and bottom panel show $R^{E}_{out}$, $R^{G}_{side}$,  $R^{E}_{long}$ radii, respectively. The data points for the multiplicity classes 20-40\% and 40-60\% have been slightly shifted in $k_{\rm T}$ for visibility.}
\label{fig:pbktcentradiiEGE}
\end{center}
\end{figure} 

The corresponding fit results for the EGE fit are shown in Fig.~\ref{fig:pbktcentradiiEGE}. In the $side$ direction the radii are consistent with the GGG results. The radii in the $out$ and $long$ directions are not Gaussian widths in this case and cannot be directly compared to previous fits. However, all the trends are qualitatively the same in both cases: radii increase with event multiplicity and decrease with pair transverse momentum. The values are 10\% (for $side$ and $long$) to 20\% (for $out$) higher than those measured in pp collisions at similar event multiplicity~\cite{Aamodt:2011kd}. The $\lambda$ parameter for the EGE fit is on the order of 0.7 for the SH fits, and growing from 0.7 at low $k_{\rm T}$ to approximately 0.9 at the highest $k_{\rm T}$ for the Cartesian fit, therefore significantly higher than in the Gaussian case. The observation that $\lambda$ is closer to unity when moving from GGG to EGE fits is expected, as the EGE fit describes the shape of the correlation much better at low $q$ and therefore better accounts for the non-Gaussian tails in the source function. 

\subsection{Model comparisons}
Hydrodynamic model calculations for p--Pb collisions~\cite{Bozek:2013df,Shapoval:2013jca}, shown as lines in Fig.~\ref{fig:pbktcentradii}, predict the existence of a collectively expanding system. Both models employ two initial transverse size assumptions: $R_{\mathrm{init}}=1.5$~fm and $R_{\mathrm{init}}=0.9$~fm, which correspond to two different scenarios of the energy deposition in the wounded nucleon model~\cite{Bozek:2013df}. The resulting charged particle multiplicity densities $\left< \mathrm{d}{N_{\mathrm{ch}}}/\mathrm{d}\eta \right>$ of 45~\cite{Bozek:2013df} and 35~\cite{Shapoval:2013jca} are equal or higher than the one in the ALICE 0-20\% multiplicity class. The calculations for $R_{\mathrm{out}}$ overestimate the measured radii, while the ones with large initial size strongly overpredict the radii. The scenarios with lower initial size are closer to the data. For $R_{\mathrm{side}}$, the calculations are in good agreement with the data in the highest multiplicity class, both in magnitude and in the slope of the $k_{\rm T}$ dependence. Only the Shapoval et al.~\cite{Shapoval:2013jca} calculation for large initial size shows higher values than data. For $R_{\mathrm{long}}$, calculations by Bo\.zek and Broniowski~\cite{Bozek:2013df} overshoot the measurement by at least 30\% for the most central data, while those by Shapoval et al. are consistent within systematics. Again, the slope of the $k_{\rm T}$ dependence is comparable. The study shows that the calculation with large initial size is disfavored by data. The calculations with lower initial size are closer to the experimental results, but are still overpredicting the overall magnitude of the radii by 10-30\%. Further refinement of the initial conditions may lead to a better agreement of the models with the data, especially at large multiplicities. The slope of the $k_{\rm T}$ dependence is usually interpreted as a signature of collectivity. Interestingly, it is very similar in data and the models in all directions, which suggests that the system dynamics might be correctly modeled by hydrodynamics.
 
Also in the data the source shape is distinctly non-Gaussian. Further studies would require examination of the source shape in p--Pb collision models to see if similar deviations from a Gaussian form are observed. 
 
The CGC approach has provided a qualitative statement on the initial size of the system in p--Pb collisions, suggesting that it is similar to that in pp~collisions \cite{Bzdak:2013zma,Schenke:2014zha}. The measured radii, at high multiplicities and low $k_{\rm T}$, are 10--20\% larger than those observed at similar multiplicities in pp~data. For lower multiplicities the differences are smaller. These differences could still be accommodated in CGC calculations. Furthermore, the evolution of the slope of the $k_{\rm T}$ dependence is similar between pp~and p--Pb collisions in the $side$ direction. Another similarity is the distinctly non-Gaussian shape of the source, which in pp and p--Pb is better described by an exponential-Gaussian-exponential form. It appears that data in p--Pb collisions still exhibit strong similarities to results from pp~collisions. However some deviations, which make the p--Pb more similar to A--A collisions, are also observed, especially at high multiplicity. The differences between small systems such as pp and p--Pb and peripheral A--A data are most naturally explained by the significantly different initial states in the two scenarios. Dedicated theoretical investigation of this issue is needed for a more definite answer, which may be able to accommodate both CGC and hydrodynamic picture.

\subsection{Comparison to the world data}
\begin{figure}[tbh!f]
\begin{center}
\includegraphics*[width=14cm]{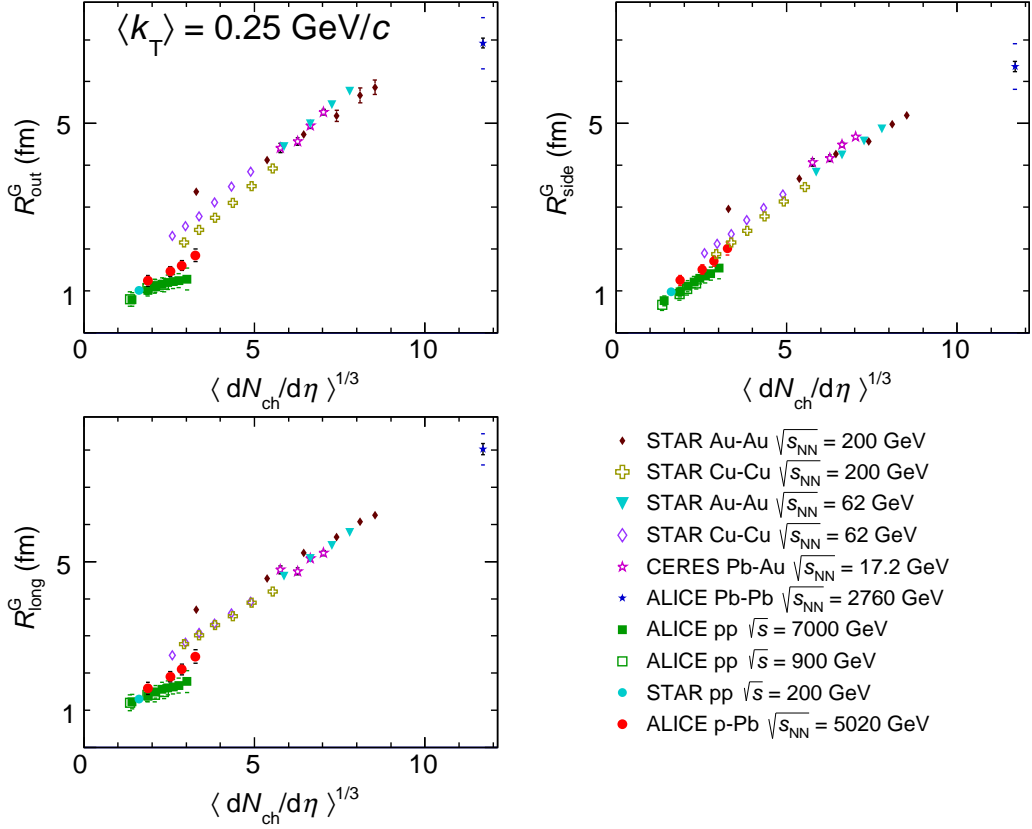} 
\caption[]{(Color online) Comparison of femtoscopic radii (Gaussian), as a function of the measured charged-particle multiplicity density, measured for various collision systems and energies by CERES~\cite{Adamova:2002wi}, STAR~\cite{Adams:2004yc,Abelev:2009tp,Aggarwal:2010aa}, PHENIX~\cite{Adler:2004rq}, and ALICE \cite{Aamodt:2011kd}.}
\label{fig:pbtopcomp}
\end{center}
\end{figure}

In Fig.~\ref{fig:pbtopcomp} the results from this analysis of the p--Pb data from the LHC (red filled circles) are compared to the world heavy-ion data, including results obtained at lower collision energies, as well as to results from pp collisions from ALICE and STAR. It has been observed~\cite{Lisa:2005dd}  that the 3-dimensional femtoscopic radii scale roughly with the cube root of the measured  charged-particle multiplicity density not only for a single energy and collision system, but also across many collision energies and initial system sizes. The pp and A--A datasets show significantly different scaling behavior, although both are linear in $\left< \mathrm{d}N_{\rm ch}/\mathrm{d}\eta \right>^{1/3}$.

The p--Pb radii agree with those in pp~collisions at low multiplicities. With increasing multiplicity, the radii for the two systems start to diverge. An analysis of one-dimensional averaged radii in pp, p--Pb and Pb--Pb collisions using the 3-pion cumulant correlations technique reveals that the multiplicity scaling for p--Pb lies between pp and Pb--Pb trends~\cite{Abelev:2014pja}, which is consistent with results presented here. On the other hand, the deviation of the correlation shape from Gaussian is similar to that observed in pp~collisions, and unlike the shapes observed in A--A collisions.

\section{Conclusions}
\label{sec:conclusions}
We reported on the three-dimensional pion femtoscopic radii in p--Pb collisions at $\sqrt{s_{\rm NN}}=5.02$~TeV, measured in four multiplicity and seven pair momentum intervals. The radii are found to decrease with $k_{\rm T}$ in all cases, similar to measurements in A--A and high-multiplicity pp~collisions. The radii increase with event multiplicity. At low multiplicities they are comparable to the pp~values, while at higher multiplicities and low pair transverse momentum they are larger by 10--20\%. However, they do not reach the values observed in A--A collisions at lower energies. The high multiplicity data are compared to predictions from two models, both of them incorporating a fast
hydrodynamic expansion of the created medium. They overpredict the values of the $R_{\mathrm{out}}$ and $R_{\mathrm{long}}$ parameters, however the introduction of a smaller initial size results in a better description. The values of the $R_{\mathrm{side}}$ parameter and the slope of the $k_{\rm T}$ dependence of the radii are in reasonable agreement. The models based on the CGC formalism suggest sizes similar to those obtained in pp~data. The observed differences of about 10--20\% for high multiplicity p--Pb collisions might not exclude this scenario. The observed non-Gaussian shape of the correlation is also similar in the pp and p--Pb collision systems. 
               %%%%%%%%%%% put the body of the article here
%
%

%%%%% acknowledgements
\newenvironment{acknowledgement}{\relax}{\relax}
\begin{acknowledgement}
\section*{Acknowledgements}
The ALICE Collaboration would like to thank all its engineers and technicians for their invaluable contributions to the construction of the experiment and the CERN accelerator teams for the outstanding performance of the LHC complex.
The ALICE Collaboration gratefully acknowledges the resources and support provided by all Grid centres and the Worldwide LHC Computing Grid (WLCG) collaboration.
The ALICE Collaboration acknowledges the following funding agencies for their support in building and
running the ALICE detector:
State Committee of Science,  World Federation of Scientists (WFS)
and Swiss Fonds Kidagan, Armenia,
Conselho Nacional de Desenvolvimento Cient\'{\i}fico e Tecnol\'{o}gico (CNPq), Financiadora de Estudos e Projetos (FINEP),
Funda\c{c}\~{a}o de Amparo \`{a} Pesquisa do Estado de S\~{a}o Paulo (FAPESP);
National Natural Science Foundation of China (NSFC), the Chinese Ministry of Education (CMOE)
and the Ministry of Science and Technology of China (MSTC);
Ministry of Education and Youth of the Czech Republic;
Danish Natural Science Research Council, the Carlsberg Foundation and the Danish National Research Foundation;
The European Research Council under the European Community's Seventh Framework Programme;
Helsinki Institute of Physics and the Academy of Finland;
French CNRS-IN2P3, the `Region Pays de Loire', `Region Alsace', `Region Auvergne' and CEA, France;
German Bundesministerium fur Bildung, Wissenschaft, Forschung und Technologie (BMBF) and the Helmholtz Association;
General Secretariat for Research and Technology, Ministry of
Development, Greece;
Hungarian Orszagos Tudomanyos Kutatasi Alappgrammok (OTKA) and National Office for Research and Technology (NKTH);
Department of Atomic Energy and Department of Science and Technology of the Government of India;
Istituto Nazionale di Fisica Nucleare (INFN) and Centro Fermi -
Museo Storico della Fisica e Centro Studi e Ricerche "Enrico
Fermi", Italy;
MEXT Grant-in-Aid for Specially Promoted Research, Ja\-pan;
Joint Institute for Nuclear Research, Dubna;
National Research Foundation of Korea (NRF);
Consejo Nacional de Cienca y Tecnologia (CONACYT), Direccion General de Asuntos del Personal Academico(DGAPA), M\'{e}xico, :Amerique Latine Formation academique – European Commission(ALFA-EC) and the EPLANET Program
(European Particle Physics Latin American Network)
Stichting voor Fundamenteel Onderzoek der Materie (FOM) and the Nederlandse Organisatie voor Wetenschappelijk Onderzoek (NWO), Netherlands;
Research Council of Norway (NFR);
National Science Centre, Poland;
Ministry of National Education/Institute for Atomic Physics and Consiliul Naţional al Cercetării Ştiinţifice - Executive Agency for Higher Education Research Development and Innovation Funding (CNCS-UEFISCDI) - Romania;
Ministry of Education and Science of Russian Federation, Russian
Academy of Sciences, Russian Federal Agency of Atomic Energy,
Russian Federal Agency for Science and Innovations and The Russian
Foundation for Basic Research;
Ministry of Education of Slovakia;
Department of Science and Technology, South Africa;
Centro de Investigaciones Energeticas, Medioambientales y Tecnologicas (CIEMAT), E-Infrastructure shared between Europe and Latin America (EELA), Ministerio de Econom\'{i}a y Competitividad (MINECO) of Spain, Xunta de Galicia (Conseller\'{\i}a de Educaci\'{o}n),
Centro de Aplicaciones Tecnológicas y Desarrollo Nuclear (CEA\-DEN), Cubaenerg\'{\i}a, Cuba, and IAEA (International Atomic Energy Agency);
Swedish Research Council (VR) and Knut $\&$ Alice Wallenberg
Foundation (KAW);
Ukraine Ministry of Education and Science;
United Kingdom Science and Technology Facilities Council (STFC);
The United States Department of Energy, the United States National
Science Foundation, the State of Texas, and the State of Ohio;
Ministry of Science, Education and Sports of Croatia and  Unity through Knowledge Fund, Croatia.
Council of Scientific and Industrial Research (CSIR), New Delhi, India
    %%%%%%% done by webmaster team
\end{acknowledgement}

%%%%%%%% Bibliography (In case of using bibtex generate the bbl requested by arXiv)
\bibliographystyle{utphys}   % Put here the style file name for the paper, i.e.apsrev4-1, utphys
\bibliography{bibliography}

%%%%%%%%% appendix with author list
\newpage
\appendix
\section{The ALICE Collaboration}
\label{app:collab}

% Collaboration: CERN-LHC-ALICE
% Generation Date is 2015/Jan/16

% How to use:
%%%%%%%%% appendix with author list
%\appendix
%\section{The ALICE Collaboration}
%\label{app:collab}
%\input{authors-list.tex}  %%%%%%% get the latest version before submitting

\begingroup
\small
\begin{flushleft}
J.~Adam\Irefn{org39}\And
D.~Adamov\'{a}\Irefn{org82}\And
M.M.~Aggarwal\Irefn{org86}\And
G.~Aglieri Rinella\Irefn{org36}\And
M.~Agnello\Irefn{org93}\textsuperscript{,}\Irefn{org110}\And
N.~Agrawal\Irefn{org47}\And
Z.~Ahammed\Irefn{org130}\And
I.~Ahmed\Irefn{org16}\And
S.U.~Ahn\Irefn{org67}\And
I.~Aimo\Irefn{org93}\textsuperscript{,}\Irefn{org110}
% ...
\renewcommand\labelenumi{\textsuperscript{\theenumi}~}

\section*{Affiliation notes}
\renewcommand\theenumi{\roman{enumi}}
\begin{Authlist}
\item \Adef{0}Deceased
\item \Adef{idp3742880}{Also at: M.V. Lomonosov Moscow State University, D.V. Skobeltsyn Institute of Nuclear Physics, Moscow, Russia}
\item \Adef{idp4017440}{Also at: University of Belgrade, Faculty of Physics and "Vin\v{c}a" Institute of Nuclear Sciences, Belgrade, Serbia}
\item \Adef{idp4354672}{Permanent Address: Permanent Address: Konkuk University, Seoul, Korea}
\item \Adef{idp4889488}{Also at: Institute of Theoretical Physics, University of Wroclaw, Wroclaw, Poland}
\item \Adef{idp5861728}{Also at: University of Kansas, Lawrence, KS, United States}
\end{Authlist}

\section*{Collaboration Institutes}
\renewcommand\theenumi{\arabic{enumi}~}
\begin{Authlist}

\item \Idef{org1}A.I. Alikhanyan National Science Laboratory (Yerevan Physics Institute) Foundation, Yerevan, Armenia
\item \Idef{org2}Benem\'{e}rita Universidad Aut\'{o}noma de Puebla, Puebla, Mexico
\item \Idef{org3}Bogolyubov Institute for Theoretical Physics, Kiev, Ukraine
\item \Idef{org4}Bose Institute, Department of Physics and Centre for Astroparticle Physics and Space Science (CAPSS), Kolkata, India
\item \Idef{org5}Budker Institute for Nuclear Physics, Novosibirsk, Russia
\item \Idef{org6}California Polytechnic State University, San Luis Obispo, CA, United States
\item \Idef{org7}Central China Normal University, Wuhan, China
\item \Idef{org8}Centre de Calcul de l'IN2P3, Villeurbanne, France
\item \Idef{org9}Centro de Aplicaciones Tecnol\'{o}gicas y Desarrollo Nuclear (CEADEN), Havana, Cuba
\item \Idef{org10}Centro de Investigaciones Energ\'{e}ticas Medioambientales y Tecnol\'{o}gicas (CIEMAT), Madrid, Spain
\item \Idef{org11}Centro de Investigaci\'{o}n y de Estudios Avanzados (CINVESTAV), Mexico City and M\'{e}rida, Mexico
\item \Idef{org12}Centro Fermi - Museo Storico della Fisica e Centro Studi e Ricerche ``Enrico Fermi'', Rome, Italy
\item \Idef{org13}Chicago State University, Chicago, USA
\item \Idef{org14}China Institute of Atomic Energy, Beijing, China
\item \Idef{org15}Commissariat \`{a} l'Energie Atomique, IRFU, Saclay, France
\item \Idef{org16}COMSATS Institute of Information Technology (CIIT), Islamabad, Pakistan
\item \Idef{org17}Departamento de F\'{\i}sica de Part\'{\i}culas and IGFAE, Universidad de Santiago de Compostela, Santiago de Compostela, Spain
\item \Idef{org18}Department of Physics and Technology, University of Bergen, Bergen, Norway
\item \Idef{org19}Department of Physics, Aligarh Muslim University, Aligarh, India
\item \Idef{org20}Department of Physics, Ohio State University, Columbus, OH, United States
\item \Idef{org21}Department of Physics, Sejong University, Seoul, South Korea
\item \Idef{org22}Department of Physics, University of Oslo, Oslo, Norway
\item \Idef{org23}Dipartimento di Elettrotecnica ed Elettronica del Politecnico, Bari, Italy
\item \Idef{org24}Dipartimento di Fisica dell'Universit\`{a} 'La Sapienza' and Sezione INFN Rome, Italy
\item \Idef{org25}Dipartimento di Fisica dell'Universit\`{a} and Sezione INFN, Cagliari, Italy
\item \Idef{org26}Dipartimento di Fisica dell'Universit\`{a} and Sezione INFN, Trieste, Italy
\item \Idef{org27}Dipartimento di Fisica dell'Universit\`{a} and Sezione INFN, Turin, Italy
\item \Idef{org28}Dipartimento di Fisica e Astronomia dell'Universit\`{a} and Sezione INFN, Bologna, Italy
\item \Idef{org29}Dipartimento di Fisica e Astronomia dell'Universit\`{a} and Sezione INFN, Catania, Italy
\item \Idef{org30}Dipartimento di Fisica e Astronomia dell'Universit\`{a} and Sezione INFN, Padova, Italy
\item \Idef{org31}Dipartimento di Fisica `E.R.~Caianiello' dell'Universit\`{a} and Gruppo Collegato INFN, Salerno, Italy
\item \Idef{org32}Dipartimento di Scienze e Innovazione Tecnologica dell'Universit\`{a} del  Piemonte Orientale and Gruppo Collegato INFN, Alessandria, Italy
\item \Idef{org33}Dipartimento Interateneo di Fisica `M.~Merlin' and Sezione INFN, Bari, Italy
\item \Idef{org34}Division of Experimental High Energy Physics, University of Lund, Lund, Sweden
\item \Idef{org35}Eberhard Karls Universit\"{a}t T\"{u}bingen, T\"{u}bingen, Germany
\item \Idef{org36}European Organization for Nuclear Research (CERN), Geneva, Switzerland
\item \Idef{org37}Faculty of Engineering, Bergen University College, Bergen, Norway
\item \Idef{org38}Faculty of Mathematics, Physics and Informatics, Comenius University, Bratislava, Slovakia
\item \Idef{org39}Faculty of Nuclear Sciences and Physical Engineering, Czech Technical University in Prague, Prague, Czech Republic
\item \Idef{org40}Faculty of Science, P.J.~\v{S}af\'{a}rik University, Ko\v{s}ice, Slovakia
\item \Idef{org41}Faculty of Technology, Buskerud and Vestfold University College, Vestfold, Norway
\item \Idef{org42}Frankfurt Institute for Advanced Studies, Johann Wolfgang Goethe-Universit\"{a}t Frankfurt, Frankfurt, Germany
\item \Idef{org43}Gangneung-Wonju National University, Gangneung, South Korea
\item \Idef{org44}Gauhati University, Department of Physics, Guwahati, India
\item \Idef{org45}Helsinki Institute of Physics (HIP), Helsinki, Finland
\item \Idef{org46}Hiroshima University, Hiroshima, Japan
\item \Idef{org47}Indian Institute of Technology Bombay (IIT), Mumbai, India
\item \Idef{org48}Indian Institute of Technology Indore, Indore (IITI), India
\item \Idef{org49}Inha University, Incheon, South Korea
\item \Idef{org50}Institut de Physique Nucl\'eaire d'Orsay (IPNO), Universit\'e Paris-Sud, CNRS-IN2P3, Orsay, France
\item \Idef{org51}Institut f\"{u}r Informatik, Johann Wolfgang Goethe-Universit\"{a}t Frankfurt, Frankfurt, Germany
\item \Idef{org52}Institut f\"{u}r Kernphysik, Johann Wolfgang Goethe-Universit\"{a}t Frankfurt, Frankfurt, Germany
\item \Idef{org53}Institut f\"{u}r Kernphysik, Westf\"{a}lische Wilhelms-Universit\"{a}t M\"{u}nster, M\"{u}nster, Germany
\item \Idef{org54}Institut Pluridisciplinaire Hubert Curien (IPHC), Universit\'{e} de Strasbourg, CNRS-IN2P3, Strasbourg, France
\item \Idef{org55}Institute for Nuclear Research, Academy of Sciences, Moscow, Russia
\item \Idef{org56}Institute for Subatomic Physics of Utrecht University, Utrecht, Netherlands
\item \Idef{org57}Institute for Theoretical and Experimental Physics, Moscow, Russia
\item \Idef{org58}Institute of Experimental Physics, Slovak Academy of Sciences, Ko\v{s}ice, Slovakia
\item \Idef{org59}Institute of Physics, Academy of Sciences of the Czech Republic, Prague, Czech Republic
\item \Idef{org60}Institute of Physics, Bhubaneswar, India
\item \Idef{org61}Institute of Space Science (ISS), Bucharest, Romania
\item \Idef{org62}Instituto de Ciencias Nucleares, Universidad Nacional Aut\'{o}noma de M\'{e}xico, Mexico City, Mexico
\item \Idef{org63}Instituto de F\'{\i}sica, Universidad Nacional Aut\'{o}noma de M\'{e}xico, Mexico City, Mexico
\item \Idef{org64}iThemba LABS, National Research Foundation, Somerset West, South Africa
\item \Idef{org65}Joint Institute for Nuclear Research (JINR), Dubna, Russia
\item \Idef{org66}Konkuk University, Seoul, South Korea
\item \Idef{org67}Korea Institute of Science and Technology Information, Daejeon, South Korea
\item \Idef{org68}KTO Karatay University, Konya, Turkey
\item \Idef{org69}Laboratoire de Physique Corpusculaire (LPC), Clermont Universit\'{e}, Universit\'{e} Blaise Pascal, CNRS--IN2P3, Clermont-Ferrand, France
\item \Idef{org70}Laboratoire de Physique Subatomique et de Cosmologie, Universit\'{e} Grenoble-Alpes, CNRS-IN2P3, Grenoble, France
\item \Idef{org71}Laboratori Nazionali di Frascati, INFN, Frascati, Italy
\item \Idef{org72}Laboratori Nazionali di Legnaro, INFN, Legnaro, Italy
\item \Idef{org73}Lawrence Berkeley National Laboratory, Berkeley, CA, United States
\item \Idef{org74}Lawrence Livermore National Laboratory, Livermore, CA, United States
\item \Idef{org75}Moscow Engineering Physics Institute, Moscow, Russia
\item \Idef{org76}National Centre for Nuclear Studies, Warsaw, Poland
\item \Idef{org77}National Institute for Physics and Nuclear Engineering, Bucharest, Romania
\item \Idef{org78}National Institute of Science Education and Research, Bhubaneswar, India
\item \Idef{org79}Niels Bohr Institute, University of Copenhagen, Copenhagen, Denmark
\item \Idef{org80}Nikhef, National Institute for Subatomic Physics, Amsterdam, Netherlands
\item \Idef{org81}Nuclear Physics Group, STFC Daresbury Laboratory, Daresbury, United Kingdom
\item \Idef{org82}Nuclear Physics Institute, Academy of Sciences of the Czech Republic, \v{R}e\v{z} u Prahy, Czech Republic
\item \Idef{org83}Oak Ridge National Laboratory, Oak Ridge, TN, United States
\item \Idef{org84}Petersburg Nuclear Physics Institute, Gatchina, Russia
\item \Idef{org85}Physics Department, Creighton University, Omaha, NE, United States
\item \Idef{org86}Physics Department, Panjab University, Chandigarh, India
\item \Idef{org87}Physics Department, University of Athens, Athens, Greece
\item \Idef{org88}Physics Department, University of Cape Town, Cape Town, South Africa
\item \Idef{org89}Physics Department, University of Jammu, Jammu, India
\item \Idef{org90}Physics Department, University of Rajasthan, Jaipur, India
\item \Idef{org91}Physik Department, Technische Universit\"{a}t M\"{u}nchen, Munich, Germany
\item \Idef{org92}Physikalisches Institut, Ruprecht-Karls-Universit\"{a}t Heidelberg, Heidelberg, Germany
\item \Idef{org93}Politecnico di Torino, Turin, Italy
\item \Idef{org94}Purdue University, West Lafayette, IN, United States
\item \Idef{org95}Pusan National University, Pusan, South Korea
\item \Idef{org96}Research Division and ExtreMe Matter Institute EMMI, GSI Helmholtzzentrum f\"ur Schwerionenforschung, Darmstadt, Germany
\item \Idef{org97}Rudjer Bo\v{s}kovi\'{c} Institute, Zagreb, Croatia
\item \Idef{org98}Russian Federal Nuclear Center (VNIIEF), Sarov, Russia
\item \Idef{org99}Russian Research Centre Kurchatov Institute, Moscow, Russia
\item \Idef{org100}Saha Institute of Nuclear Physics, Kolkata, India
\item \Idef{org101}School of Physics and Astronomy, University of Birmingham, Birmingham, United Kingdom
\item \Idef{org102}Secci\'{o}n F\'{\i}sica, Departamento de Ciencias, Pontificia Universidad Cat\'{o}lica del Per\'{u}, Lima, Peru
\item \Idef{org103}Sezione INFN, Bari, Italy
\item \Idef{org104}Sezione INFN, Bologna, Italy
\item \Idef{org105}Sezione INFN, Cagliari, Italy
\item \Idef{org106}Sezione INFN, Catania, Italy
\item \Idef{org107}Sezione INFN, Padova, Italy
\item \Idef{org108}Sezione INFN, Rome, Italy
\item \Idef{org109}Sezione INFN, Trieste, Italy
\item \Idef{org110}Sezione INFN, Turin, Italy
\item \Idef{org111}SSC IHEP of NRC Kurchatov institute, Protvino, Russia
\item \Idef{org112}SUBATECH, Ecole des Mines de Nantes, Universit\'{e} de Nantes, CNRS-IN2P3, Nantes, France
\item \Idef{org113}Suranaree University of Technology, Nakhon Ratchasima, Thailand
\item \Idef{org114}Technical University of Split FESB, Split, Croatia
\item \Idef{org115}The Henryk Niewodniczanski Institute of Nuclear Physics, Polish Academy of Sciences, Cracow, Poland
\item \Idef{org116}The University of Texas at Austin, Physics Department, Austin, TX, USA
\item \Idef{org117}Universidad Aut\'{o}noma de Sinaloa, Culiac\'{a}n, Mexico
\item \Idef{org118}Universidade de S\~{a}o Paulo (USP), S\~{a}o Paulo, Brazil
\item \Idef{org119}Universidade Estadual de Campinas (UNICAMP), Campinas, Brazil
\item \Idef{org120}University of Houston, Houston, TX, United States
\item \Idef{org121}University of Jyv\"{a}skyl\"{a}, Jyv\"{a}skyl\"{a}, Finland
\item \Idef{org122}University of Liverpool, Liverpool, United Kingdom
\item \Idef{org123}University of Tennessee, Knoxville, TN, United States
\item \Idef{org124}University of the Witwatersrand, Johannesburg, South Africa
\item \Idef{org125}University of Tokyo, Tokyo, Japan
\item \Idef{org126}University of Tsukuba, Tsukuba, Japan
\item \Idef{org127}University of Zagreb, Zagreb, Croatia
\item \Idef{org128}Universit\'{e} de Lyon, Universit\'{e} Lyon 1, CNRS/IN2P3, IPN-Lyon, Villeurbanne, France
\item \Idef{org129}V.~Fock Institute for Physics, St. Petersburg State University, St. Petersburg, Russia
\item \Idef{org130}Variable Energy Cyclotron Centre, Kolkata, India
\item \Idef{org131}Warsaw University of Technology, Warsaw, Poland
\item \Idef{org132}Wayne State University, Detroit, MI, United States
\item \Idef{org133}Wigner Research Centre for Physics, Hungarian Academy of Sciences, Budapest, Hungary
\item \Idef{org134}Yale University, New Haven, CT, United States
\item \Idef{org135}Yonsei University, Seoul, South Korea
\item \Idef{org136}Zentrum f\"{u}r Technologietransfer und Telekommunikation (ZTT), Fachhochschule Worms, Worms, Germany
\end{Authlist}
\endgroup

  %%%%%%% done by webmaster team
\end{document}